\documentclass{ws-mpla}

\usepackage{graphicx} \usepackage{epsfig} \usepackage{amsmath,amssymb}

\usepackage[dvips]{color}

\def\esym{$E_{sym}(\rho)$~}

\def\es0{$E_{sym}(\rho_0)$~}

\def\us0{$U_{sym}(\rho_0,k_F)$~}

\def\lr{$L(\rho)$~}

\def\l0{$L(\rho_0)$~}

\begin{document}

\markboth{Neutron-proton effective mass splitting} {Bao-An Li \& Lie-Wen Chen}

\catchline{}{}{}{}{}

\title{Neutron-proton effective mass splitting in neutron-rich matter and its impacts on nuclear reactions}


\author{\footnotesize Bao-An Li} \address{Department of Physics \&
  Astronomy, Texas A\&M University-Commerce \\ Commerce, Texas 75429,
  USA\\Bao-An.Li@tamuc.edu}

\author{\footnotesize Lie-Wen Chen}
\address{Department of Physics and Astronomy and Shanghai Key Laboratory for Particle Physics and Cosmology, Shanghai Jiao Tong University, Shanghai 200240, China\\Lwchen@sjtu.edu.cn}

\maketitle

\pub{Received (Day Month Year)}{Revised (Day Month Year)}

\begin{abstract}
The neutron-proton effective mass splitting in neutron-rich nucleonic matter reflects the space-time nonlocality of the isovector nuclear interaction. It affects the neutron/proton ratio during the earlier evolution of the Universe, cooling of protoneutron stars, structure of rare isotopes and dynamics of heavy-ion collisions. While there is still no consensus on whether the neutron-proton effective mass splitting is negative, zero or positive and how it depends on the density as well as the isospin-asymmetry of the medium, significant progress has been made in recent yeas in addressing these issues. There are different kinds of nucleon effective masses. In this mini-review, we focus on the total effective masses often used in the non-relativistic description of nuclear dynamics. We first  recall the connections among the neutron-proton effective mass splitting, the momentum dependence of the isovector potential and the density dependence of the symmetry energy. We then make  a few observations about the progress in calculating the
neutron-proton effective mass splitting using various nuclear many-body theories and its effects on the isospin-dependence of in-medium nucleon-nucleon cross sections. Perhaps, our most reliable knowledge
so far about the neutron-proton effective mass splitting at saturation density of nuclear matter comes from optical model analyses of huge sets of nucleon-nucleus scattering data accumulated over the last five decades. The momentum dependence of the symmetry potential from these analyses provide a useful boundary condition at saturation density for calibrating nuclear many-body calculations.  Several observables in heavy-ion collisions have been identified as sensitive probes of the neutron-proton effective mass splitting in dense neutron-rich matter based on transport model simulations. We review these observables and comment on the latest experimental findings.
\end{abstract}

\ccode{PACS Nos.: 21.65.Ef, 24.10.Ht, 21.65.Cd}

\section{Introduction}
At the outset, we would like to first emphasize that there are many comprehensive research and review articles on nucleon effective masses in nuclear medium in the literature.  While there are many
interesting research articles on the neutron-proton effective mass splitting or nucleon isovector effective mass in isospin-asymmetric nuclear matter, to our best knowledge few review articles exist. This brief review on the neutron-proton effective mass splitting is based mainly on our recent work originally published in refs. \cite{Li04,LiBA04,LiChen05,zuo05,XuC10,Rchen,LiXH13,LiXH14,LiBA13,Jxu15a} in collaboration with several colleagues. While we try to be inclusive and minimize our probably unavoidable biases in making some observations and comments on several key issues, this article is not intended to be a comprehensive review of the field. Moreover, there are several different kinds of nucleon effective masses, see, e.g., refs.~\cite{Jeu76,Mah85,jamo} for very extensive reviews. In particular, the concept of the nucleon effective mass in the relativistic case is much more complicated and one must be very careful in comparing the effective masses used in relativistic and non-relativistic models as stressed in refs.~\cite{Fuc05,Ron06,Che07}. In this article, we shall concentrate on the total nucleon effective mass and especially the corresponding neutron-proton effective mass splitting $m^*_{n-p}\equiv(m_{\rm n}^*-m_{\rm p}^*)/m$ used typically in the non-relativistic description of nuclear dynamics.

As it is well known, the nucleon effective mass is a fundamental quantity characterizing the propagation of a nucleon in nuclear medium \cite{Jeu76,jamo,sjo76,Neg81}. It describes to leading order the effects related to the space-time non-locality of the underlying nuclear effective interactions and the Pauli exchange effects in nuclear systems. It is also directly related to the discontinuity of momentum distribution of correlated nucleons at the Fermi surface \cite{Jeu76,Mig57,Lut60,Czy61,Sar80,Bla81,Kro81,Jac82}. In isospin-asymmetric nucleonic matter, whether the effective mass for neutrons is higher, equal to or lower than that for protons and how their difference may depend on properties of the medium are among the interesting new questions arising. As summarized nicely by Mei$\beta$ner {\it et al.} \cite{Mei07}, answers to these questions have important ramifications in addressing many unresolved issues in nuclear physics, astrophysics and cosmology. For instance, it is relevant for the primordial nucleosynthesis where the equilibrium neutron/proton ratio is determined by $(n/p)_{eq}=e^{-m^*_{n-p}/T}$ in the early ($\geq$ 1ms) universe when the temperature $T$ was high ($\geq$ 3 MeV) \cite{Ste06}. It is also important for calculating the neutrino opacities in neutron star matter \cite{Bur06}. Moreover, it is critical for understanding properties of mirror nuclei \cite{Nol69}, the location of drip-lines \cite{Wod97}, energy level densities of reaction partners \cite{Cha05}, the time scale and degree of isospin transport in nuclear reactions induced by radioactive beams \cite{LiBA04,Riz05}, isospin-sensitive observables in heavy-ion collisions \cite{LiBA04,LiChen05,Riz05,Gio10,Feng12,Zhang14,Xie14} as well as thermal and transport properties of neutron-rich matter \cite{Jxu15a,Beh11,Jxu15b}. Furthermore, we notice here that the individual effective masses of neutrons and protons enter into many microphysics processes in both nuclear physics and astrophysics. For example, in different cooling processes of proto-neutron stars, the product of various powers of $(m_n^*/m_n)^i\times (m_p^*/m_p)^j$ are involved in calculating the neutrino emission rates. The different effective masses of neutrons and protons, although not always directly their difference, play a significantly role in determining several properties of neutron stars \cite{Yak01,Pag06,Bal14}. In addition, it has long been known that the nucleon effective masses affect the in-medium nucleon-nucleon cross sections, thus transport and thermal properties of nuclear matter.

Besides continuing efforts in calculating the neutron-proton effective mass splitting $m^{*}_{n-p}$ using essentially all available microscopic many-body theories and phenomenological approaches with various interactions, significant efforts were devoted recently to extracting the $m^{*}_{n-p}$ from nucleon-nucleus scattering experiments \cite{XuC10,LiXH14}.  Moreover, several experimental observables in heavy-ion collisions sensitive to the neutron-proton effective mass splitting were predicted based on transport model simulations \cite{LiBA04,LiChen05,Riz05,Gio10,Feng12,Zhang14,Xie14}. Interestingly, some indications about the sign of the $m^{*}_{n-p}$ in neutron-rich matter from transport model analyses of heavy-ion reaction data were also reported very recently \cite{Cou14}. In this brief review, we summarize a few key progresses in these efforts in order to stimulate further studies on this currently hotly debated topic, especially in view of the great new opportunities of studying properties of neutron-rich matter at rare isotope beam facilities under construction around the world.

\section{Neutron-proton effective mass splitting and momentum dependence of symmetry (isovector) potential}
Microscopic nuclear many-body theories indicate that the real part of the single-nucleon potential  $U_{\tau}(k, \mathcal{E},\rho,\delta)$ for $\tau=n$ or $p$ in nuclear matter of density $\rho$ and isospin-asymmetry $\delta\equiv (\rho_n-\rho_p)/\rho$ depends on not only the nucleon momentum $k$ but also its energy $\mathcal{E}$, reflecting the nonlocality in both space and time of nuclear interactions, see, e.g. \cite{jamo,LLL} for detailed discussions. These two kinds of nonlocality can be characterized by using the so-called nucleon effective  k-mass and E-mass, respectively defined in terms of the partial derivative of $U$ with respect to $k$ and $\mathcal{E}$ \cite{jamo}. However, once a dispersion relation
$k(\mathcal{E})$ or $\mathcal{E}(k)$ is known from the on-shell condition $\mathcal{E}=k^2/2m+U(k, \mathcal{E},\rho,\delta)$, an equivalent potential either local in space or time, i.e.,
$U(k(\mathcal{E}), \mathcal{E},\rho,\delta)$ or $U(k,\mathcal{E}(k),\rho,\delta)$, can be obtained.  Thus, the total nucleon effective mass can be calculated using either the first or second part of its defining equation depending on whether the $\mathcal{E}$ or $k$ is selected as the explicit variable \cite{jamo}
\begin{equation}\label{em1}
\frac{m^{*}_{\tau}}{m_{\tau}}=1-\frac{dU_{\tau}(k(\mathcal{E}),\mathcal{E},\rho,\delta)}{d\mathcal{E}}
=\left[1+\frac{m_{\tau}}{\hbar^2k_F^{\tau}}\frac{dU_{\tau}(k,\mathcal{E}(k),\rho,\delta)}{dk}\Bigg|_{k_F^{\tau}}\right]^{-1}
\end{equation}
where $m_{\tau}$ represents the mass of neutrons or protons in free-space and the neutron/proton Fermi momentum $k_F^{\tau}=(1+\tau_3\delta)^{1/3}\cdot k_F$ with $k_F=(3\pi^2\rho/2)^{1/3}$  being the nucleon Fermi momentum in symmetric matter at density $\rho$ and $\tau_3=+1$ or $-1$ for neutrons or protons.

It is well known that the nucleon potential $U_{\tau}(k,\rho,\delta)$ in isospin-asymmetric matter can be expanded in $\delta$ as
\begin{equation}\label{sp}
U_{\tau}(k,\rho,\delta)=U_0(k,\rho)+\tau_3 U_{sym,1}(k,\rho)\cdot\delta+U_{sym,2}(k,\rho)\cdot\delta^2+\tau_3\mathcal{O}(\delta^3),
\end{equation}
where $U_0(k,\rho)$, $U_{sym,1}(k,\rho)$ (conventionally denoted as $U_{sym}(k,\rho)$, namely, the Lane potential~\cite{Lan62}) and $U_{sym,2}(k,\rho)$ are the isoscalar, isovector (symmetry) and second-order isoscalar potentials, respectively.
The neutron-proton effective mass splitting $m^*_{n-p}(\rho,\delta)$ is then
\begin{align}\label{em2}
m^*_{n-p}=\frac{\frac{m}{\hbar^2}\left(\frac{1}{k_F^p}\frac{dU_p}{dk}\mid_{k_F^p}-\frac{1}{k_F^n}\frac{dU_n}{dk}\mid_{k_F^n}\right)}{\left[1+\frac{m_p}{\hbar^2k_F^p}\frac{dU_p}{dk}\mid_{k_F^p}\right]\left[1+\frac{m_n}{\hbar^2k_F^n}\frac{dU_n}{dk}\mid_{k_F^n}\right]}.
\end{align}
Considering that the $U_{sym,1}(\rho,k)\cdot\delta$ term is always much smaller than the isoscalar potential $U_0(\rho,k)$ in Eq.~(\ref{sp}), it was shown~\cite{LiBA13} that the denominator in Eq.~(\ref{em2}) can be well approximated by $(1+\frac{m}{\hbar^2k_{\rm F}}dU_p/dk)(1+\frac{m}{\hbar^2k_{\rm F}}dU_n/dk)\approx (1+\frac{m}{\hbar^2k_{\rm F}}dU_0/dk)^2=(m/m^*_0)^2$. Up to the first-order in isospin asymmetry parameter $\delta$, the expression for $m^*_{n-p}(\rho,\delta)$ can be further simplified to
\begin{equation}\label{npe1}
m^*_{n-p}\approx 2\delta\frac{m}{\hbar^2k_F}\left[-\frac{dU_{sym,1}}{dk}-\frac{k_F}{3}\frac{d^2U_0}{dk^2}+\frac{1}{3}\frac{dU_0}{dk}\right]_{k_F}\left(\frac{m^*_0}{m}\right)^2.
\end{equation}
This expression is valid at an arbitrary density $\rho$ and it is seen that the $m^*_{n-p}$ generally depends explicitly on the momentum dependence of  both the isovector $U_{sym,1}$ and
isoscalar $U_0$ potentials. At saturation density $\rho_0$, isospin-dependent optical model analyses of nucleon-nucleus scattering data \cite{LiXH14} indicate that the last two terms in Eq. (\ref{npe1}) largely cancel out, leaving the momentum dependence of the symmetry potential $dU_{sym,1}/dk$ as the dominating factor. Much information about the momentum dependence of the isoscalar potential $dU_0/dk$ especially at saturation density and the associated nucleon isoscalar effective mass $m^*_0$ at the corresponding Fermi surface is available. However, our current knowledge about the momentum dependence of the isovector potential is still rather poor. In fact, as we shall summarize in the next section, different microscopic and/or phenomenological many-body theories using various nuclear interactions predict rather diverse momentum dependences for the isovector potential. Consequently, the predicted neutron-proton effective mass splitting in isospin-asymmetric nucleonic matter is still very uncertain.

\begin{figure}[htb]
\begin{center}
\includegraphics[width=13.cm,height=6.cm]{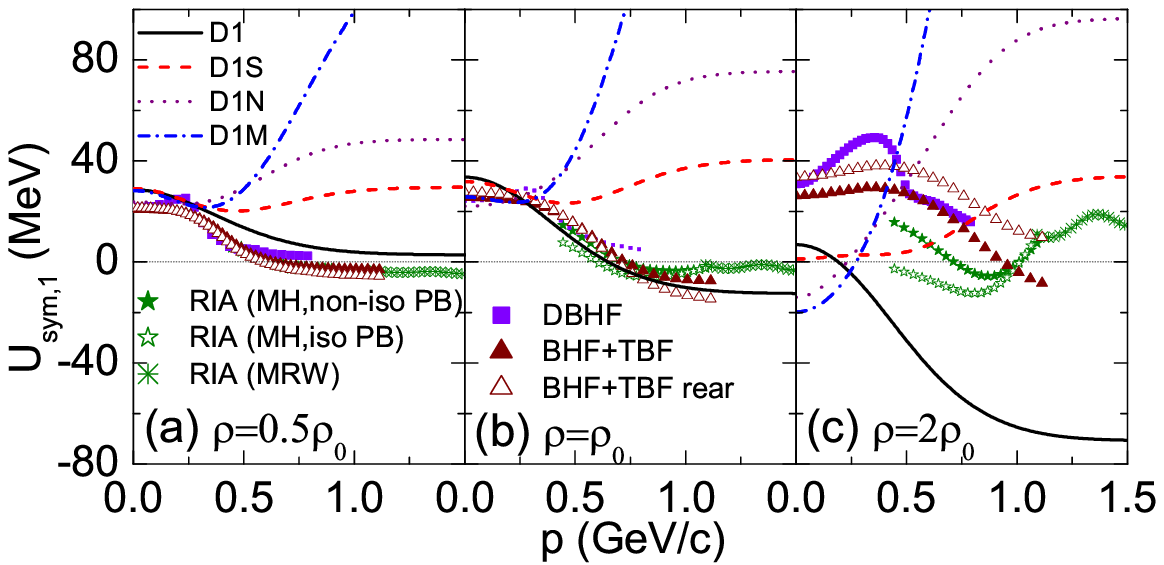}
\includegraphics[width=13.cm,height=6.cm]{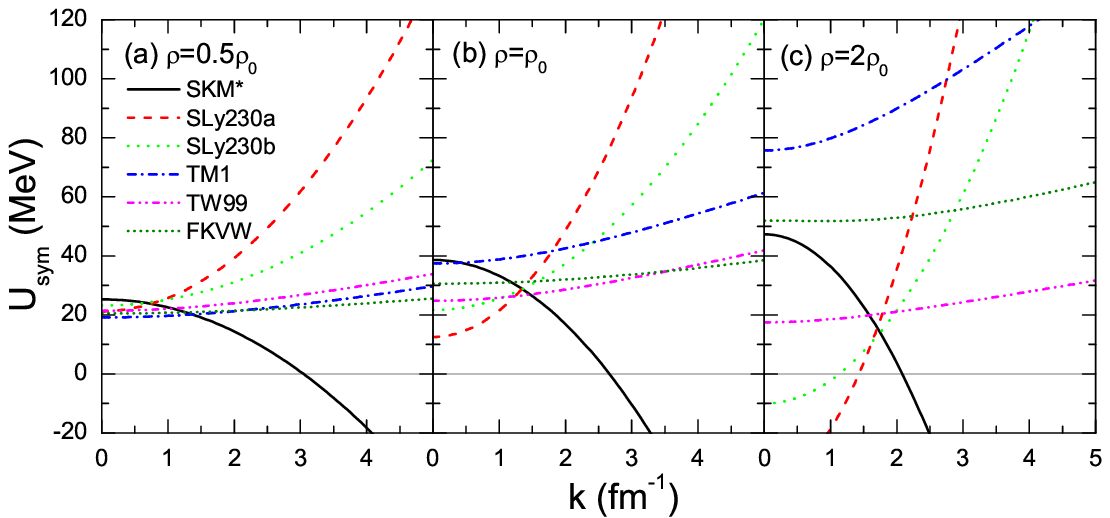}
\caption{((Upper windows) Density and momentum dependence of the nucleon isovector potential predicted by the Gogny-Hartree-Fock calculations using the D1, D1S, D1M and D1N interactions, Dirac-Brueckner-Hartree-Fock (DBHF) and Relativistic Impulse Approximation (RIA) with various two-body and three-body forces (TBF). Taken from ref. \protect\cite{Rchen}. (Low windows) Predictions using the SHF and RMF approaches at a density of $\protect\rho= 0.5\protect\rho_{0}$, $1.0\protect\rho _{0}$ and $2\protect\rho _{0}$, respectively. Taken from ref. \protect\cite{LCK08} }\label{Usym1}
\end{center}
\end{figure}
\section{Physics origins of the uncertainties in the neutron-proton effective mass splitting in neutron-rich matter}
The nuclear symmetry potential has been calculated using various microscopic theories and phenomenological approaches, such as the
relativistic DBHF theory \cite{Fuc05,Ron06,Fuc04,Ma04,Sam05a,Ulr97}, the non-relativistic BHF theory \cite{zuo05,Bom91}, the chiral
perturbation theory \cite{Fri05}, the RMF approach \cite{Che07,Bar05}, the non-relativistic mean-field models based on Skyrme-like interactions \cite{Li04,Das03,Beh05,Xu07c}, and the relativistic impulse approximation \cite{Che05c,LiZH06b}. Unfortunately, they give widely different predictions for the momentum dependence of the nuclear symmetry potential especially at high densities and/or momenta.
Shown as examples in the upper windows of Fig.~\ref{Usym1} are the symmetry potentials predicted using the Gogny-Hartree-Fock, Dirac-Brueckner-Hartree-Fock and the Relativistic Impulse Approximation with various two-body and three-body interactions \cite{Rchen}. The lower windows are example of the SHF and RMF model predictions~\cite{LCK08} at a density of $\protect\rho= 0.5\protect\rho_{0}$, $1.0\protect\rho _{0}$ and $2\protect\rho _{0}$, respectively. It is seen that while some models predict decreasing symmetry potentials (leading to a positive $m^{*}_{n-p}$) albeit at different rates, some others predict instead increasing ones with growing nucleon momentum especially at suprasaturation densities. For instance, the four widely used Gogny interactions D1~\cite{Gogny}, D1S~\cite{Ber91}, D1N~\cite{Cha08} and D1M~\cite{Gor09} predict very different momentum dependence for the isovector potential. Thus, they predict also very different neutron-proton effective mass splittings especially at high densities and/or momenta.

Why is the isovector potential so uncertain especially at high-densities and/or high-momenta? Besides the obviously different techniques used in treating quantum many-body problems,  there are some physics origins. Some useful hints can be obtained from the expression of the isovector potential at $k_F$ in the interacting Fermi gas model \cite{pre,Xu-tensor} $ U_{sym,1}(k_F,\rho)=
\frac{1}{4}\rho\int [V_{T1}(r_{ij})f^{T1}(r_{ij})-V_{T0}(r_{ij})f^{T0}(r_{ij})]d^3r_{ij}
$ in terms of the isosinglet (T=0) and isotriplet (T=1)
nucleon-nucleon interactions $V_{T0}(r_{ij})$ and
$V_{T1}(r_{ij})$, and the corresponding nucleon-nucleon correlation functions
$f^{T0}(r_{ij})$ and $f^{T1}(r_{ij})$, respectively. Needless to
say, if there is no isospin dependence in both the nucleon-nucleon interaction
and the correlation function, then the isovector potential $
U_{sym,1}(k_F,\rho)$ vanishes. The $E_{sym}(\rho)$ thus reflects the
competition of the nucleon-nucleon interaction strengths and correlation
functions between the isosinglet and isotriplet channels.
Among the key factors affecting the competition are (1) the spin-isospin dependence of the three-body force (which is normally represented by a density-dependent effective two-body force after integrating over the third nucleon), (2) tensor forces mostly in the isosinglet channel and (3) the isospin dependence of nucleon-nucleon correlations \cite{Xu-tensor}. Our poor knowledge about the in-medium properties of these factors, such as the uncertain short-range behavior of the tensor force due to $\rho$-meson exchange, contribute dominantly to the uncertain density and momentum dependence of the isovector potential especially at supra-saturation densities \cite{Xu-tensor,Wang12}.

\section{Neutron-proton effective mass splitting and density dependence of nuclear symmetry energy}
Nuclear symmetry energy \esym encodes the energy associated with the neutron-proton asymmetry in the Equation of State (EOS) of isospin-asymmetric matter. It is currently the most uncertain part of the EOS of neutron-rich nucleonic matter especially at supra-saturation densities. Using the Hugenholtz-Van Hove (HVH) theorem \cite{hug} or the Bruckner theory \cite{bru64,Dab73}, nuclear symmetry energy \esym and its density slope $L(\rho) \equiv \left[3 \rho (\partial E_{\rm sym}/\partial \rho\right]_{\rho}$
can be expressed as \cite{XuC10,Rchen,xuli2}
\begin{eqnarray}
&& E_{\rm sym}(\rho) = \frac{1}{3} \frac{\hbar^2 k_F^2}{2 m_0^*} +
\frac{1}{2} U_{\rm sym,1}(\rho,k_{F}), \label{Esymexp2}
\\
&& L(\rho) = \frac{2}{3} \frac{\hbar^2 k_F^2}{2 m_0^*} - \frac{1}{6}\Big(\frac{\hbar^2 k^3}{{m_0^*}^2}\frac{\partial m_0^*}{\partial k} \Big)|_{k_F} + \frac{3}{2} U_{\rm sym,1}(\rho,k_F) + \frac{dU_{\rm sym,1}}{dk}|_{k_F} k_F + 3U_{\rm sym,2}(\rho,k_F). \notag\\ \label{Lexp2}
\end{eqnarray}
\begin{figure}[htb]
\begin{center}
\includegraphics[width=13cm,height=7cm]{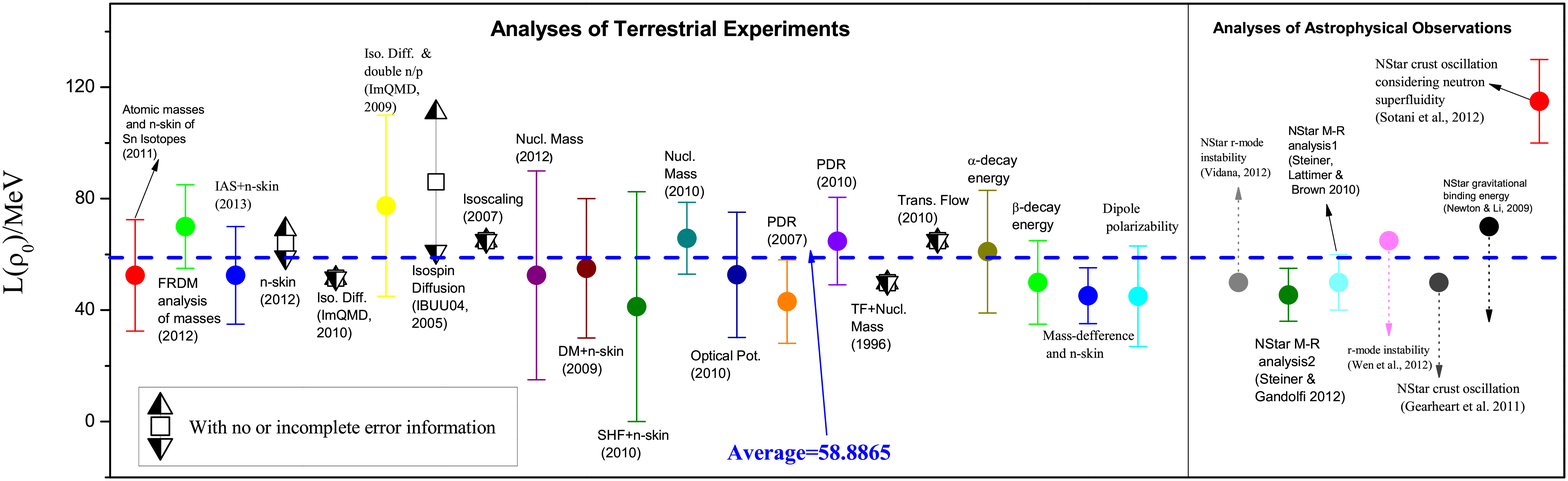}
\caption{The density slope L of nuclear symmetry energy at normal density from 28 analyses of terrestrial nuclear laboratory experiments and astrophysical observations. Taken from ref. \protect\cite{LiBA13}.}\label{L0fig}
\end{center}
\end{figure}
Comparing these expressions with Eq. (\ref{npe1}), it is seen that the \esym, \lr and $m^*_{n-p}$ depend on some common quantities especially the momentum dependence of the symmetry potential $dU_{\rm sym,1}/dk$. Thus, the knowledge about the $m^*_{n-p}$ may help us pin down the uncertain \esym and its density slope \lr or vice versa. At $\rho_0$, by neglecting the momentum dependence of the $m_0^*$ and the higher-order term $U_{\rm sym,2}(\rho,k_F)$, the Eq. (\ref{npe1}) can be further simplified in terms of the characteristics of the density dependence of nuclear symmetry energy as \cite{LiBA13}
\begin{equation}
m^*_{n-p}(\rho_0,\delta)\approx\delta\cdot \left[3E_{\rm sym}(\rho_0)-L(\rho_0)-\frac{1}{3}\frac{m}{m^*_0}E_{F}(\rho_0)\right]\bigg/ \left[E_F(\rho_0)\cdot (m/m_0^*)^2\right]. \label{npemass2}
\end{equation}
It is clear that whether the $m_{\rm n}^*$ is equal, larger or smaller than the $m_{\rm p}^*$ depends on the value of $L(\rho_0)$ relative to the quantity $[3E_{\rm sym}(\rho_0)-\frac{1}{3}\frac{m}{m^*_0}E_{F}(\rho_0)]$. For example, using relatively well determined values of \es0=31 MeV, $m_0^*/m=0.7$ and $E_F(\rho_0)=36$ MeV, a positive value for the $m^*_{n-p}(\rho_0,\delta)$ requires that \l0$\leq 76$ MeV. Indeed, as shown in Fig. \ref{L0fig}, most of the 28 studies of various terrestrial nuclear laboratory experiments and astrophysical observations have found \l0 values to be less than 76 MeV. In fact, the 2013 global average of \l0 is $58.9\pm 16$ MeV. Thus, the available information on the symmetry energy strongly indicates that the $m^*_{n-p}$ is positive \cite{LiBA13}.

\begin{figure}
\centerline{\includegraphics[width=9cm]{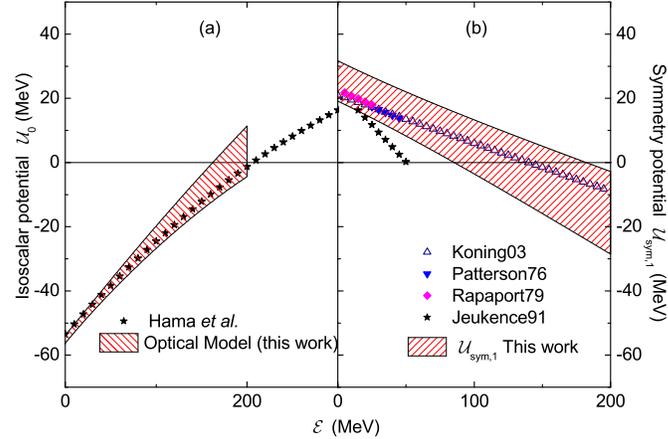}}
\label{Uoptical}
\caption{Energy dependent isoscalar $\mathcal{U}_0$(left) and isovector $\mathcal{U}_{sym,1}$ (right) nucleon potentials from analyzing nucleon-nucleus scattering data. Taken from \protect\cite{LiXH14}.}
\end{figure}
\section{Neutron-proton effective mass splitting at saturation density from optical model analyses of nucleon-nucleus scattering data}
Nucleon-nucleus scattering experiments have long been the main source of information about the energy/momentum and  isospin dependence of the nucleon optical potential at saturation density since the earlier 1960s \cite{Hod94}. While many analyses extracted only limited information about the isovector potential in segmented energy ranges up to about 200 MeV often with very poor or unknown uncertainties, they do indicate consistently that the $m^{*}_{n-p}$ is positive, see, e.g., refs.~\cite{XuC10,Cha14}. Interestingly, the very recent global optical model analysis \cite{LiXH14} of all 2249 data sets of reaction and angular differential cross sections of neutron and proton scattering on 234 targets at beam energies from 0.05 to 200 MeV available in the EXFOR database at the Brookhaven National Laboratory~\cite{Exfor} has extracted a value of $m^{*}_{n-p}=(0.41\pm0.15)\delta$. To our best knowledge, this is currently the most stringent and reliable constraint on the neutron-proton effective mass splitting at normal density. To be more specific and emphasize a few key points, we summarize in the following the most important results from ref. \cite{LiXH14}. Shown in the left window of Fig.\ \ref{Uoptical} is a comparison of the nucleon isoscalar $\mathcal{U}_0$ potentials from the non-relativistic analysis (hatched bands) \cite{LiXH14} and the Schr$\ddot{\mathrm{o}}$dinger equivalent isoscalar potential obtained by Hama \textit{et al.} using the Dirac phenomenology~\cite{Ham90}.  They are consistent up to about 200 MeV of nucleon energy with a slope leading to an
isoscalar effective mass of $m^{*}_0/m=0.65\pm 0.06$ consistent with the empirical values from many other analyses~\cite{jamo,Jeu76}.
Shown in the right window are the energy dependence of the nucleon isovector potential $\mathcal{U}_{sym,1}$ from several earlier studies~\cite{Kon03,Jeu91,Rap79,Pat76} and the most comprehensive one done very recently (hatched bands) \cite{LiXH14}.  Most of the earlier parameterizations for the $\mathcal{U}_{sym,1}$ are valid in different energy ranges while the ones by Koning \textit{et al.}~\cite{Kon03} and Li \textit{et al.}~\cite{LiXH14} are valid up to 200 MeV. Except the one by Jeukenne \textit{et al.}~\cite{Jeu91}, most of the results fall within the error band and they all clearly indicate a decreasing isovector optical potential with increasing energy.
As emphasized in refs. \cite{LiXH13,LiXH14,Dab64}, because of the isospin-dependent nucleon dispersion relation the single-nucleon potential  $U_0(\rho_0,k)$ and $U_{sym,1}(\rho_0,k)$ in isospin-asymmetric nuclear matter are related to the nucleon optical potential $\mathcal{U}_0(\mathcal{E})$ and $\mathcal{U}_{sym,1}(\mathcal{E})$  according to
\begin{equation}\label{Tran}
U_0(T(\mathcal{E}))=\mathcal{U}_0(\mathcal{E}),\,\,\,\,\,U_{sym,1}(T(\mathcal{E}))=\frac{\mathcal{U}_{sym,1}}{\mu}
\end{equation}
where $\mu=1-\frac{\partial \mathcal{U}_0}{\partial \mathcal{E}}$, $\mathcal{E}=T+U_0(T)$ and T is the nucleon kinetic energy.
\begin{figure}
\centering
\includegraphics[width=9cm]{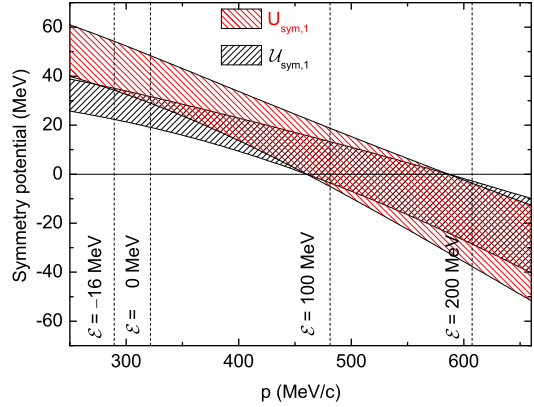}
\caption{Momentum dependence of the symmetry potential in the nucleon optical potential $\mathcal{U}_{sym,1}$ (black) and nuclear matter $U_{sym,1}$ (red), respectively. Taken from \protect\cite{LiXH14}.}
\label{U11}
\end{figure}
Shown in Fig.\ \ref{U11} is a comparison of the symmetry potential $\mathcal{U}_{sym,1}$ in the nucleon optical potential and the $U_{sym,1}$  in isospin-asymmetric nucleonic matter as a function of nucleon momentum. It is seen that their slopes are significantly different especially around the nucleon Fermi momentum of 270 MeV/c.
The momentum dependence of the $U_{sym,1}$ provides a significant boundary condition for the isovector potentials used in transport model simulations of heavy-ion reactions especially those induced by rare
isotopes. Moreover, the $U_{sym,1}$ instead of the  $\mathcal{U}_{sym,1}$ should be used in evaluating the neutron-proton effective mass splitting. To evaluate the $m^{*}_{n-p}$ according to
Eq.~(\ref{npe1}), one also needs to know not only the first-order derivative of the symmetry potential but also both the first-order and second-order derivatives of the isoscalar potential $U_0$ with respect to nucleon momentum. Their values were found to be $-dU_{sym,1}/dk=30.44$, $-k_F/3d^2U_0/dk^2=-12.88$, and $1/3dU_0/dk=10.17$ at $k_F$. Thus, the last two terms due to the momentum dependence of the isoscalar potential largely cancel out, leaving the momentum dependence of the isovector potential $-dU_{sym,1}/dk$ as the dominating source of the $m^{*}_{n-p}=(0.41\pm0.15)\delta$ at saturation
density \cite{LiXH14}.

\section{Effects of neutron-proton effective mass splitting on the isospin-dependence of in-medium nucleon-nucleon cross sections}

\begin{figure}[tbh]
\includegraphics[height=0.3\textheight,angle=-90]{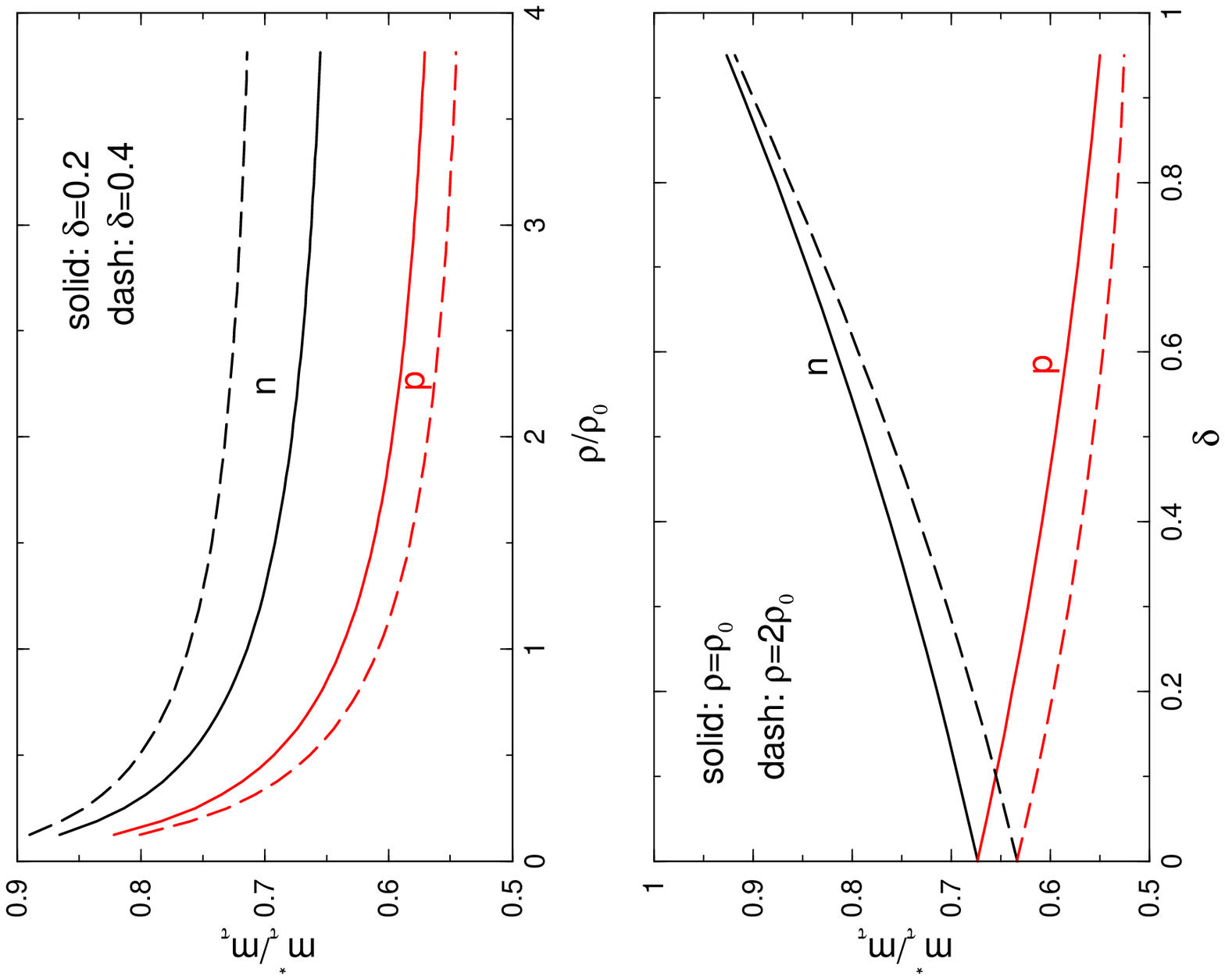}
\includegraphics[height=0.32\textheight,angle=-90]{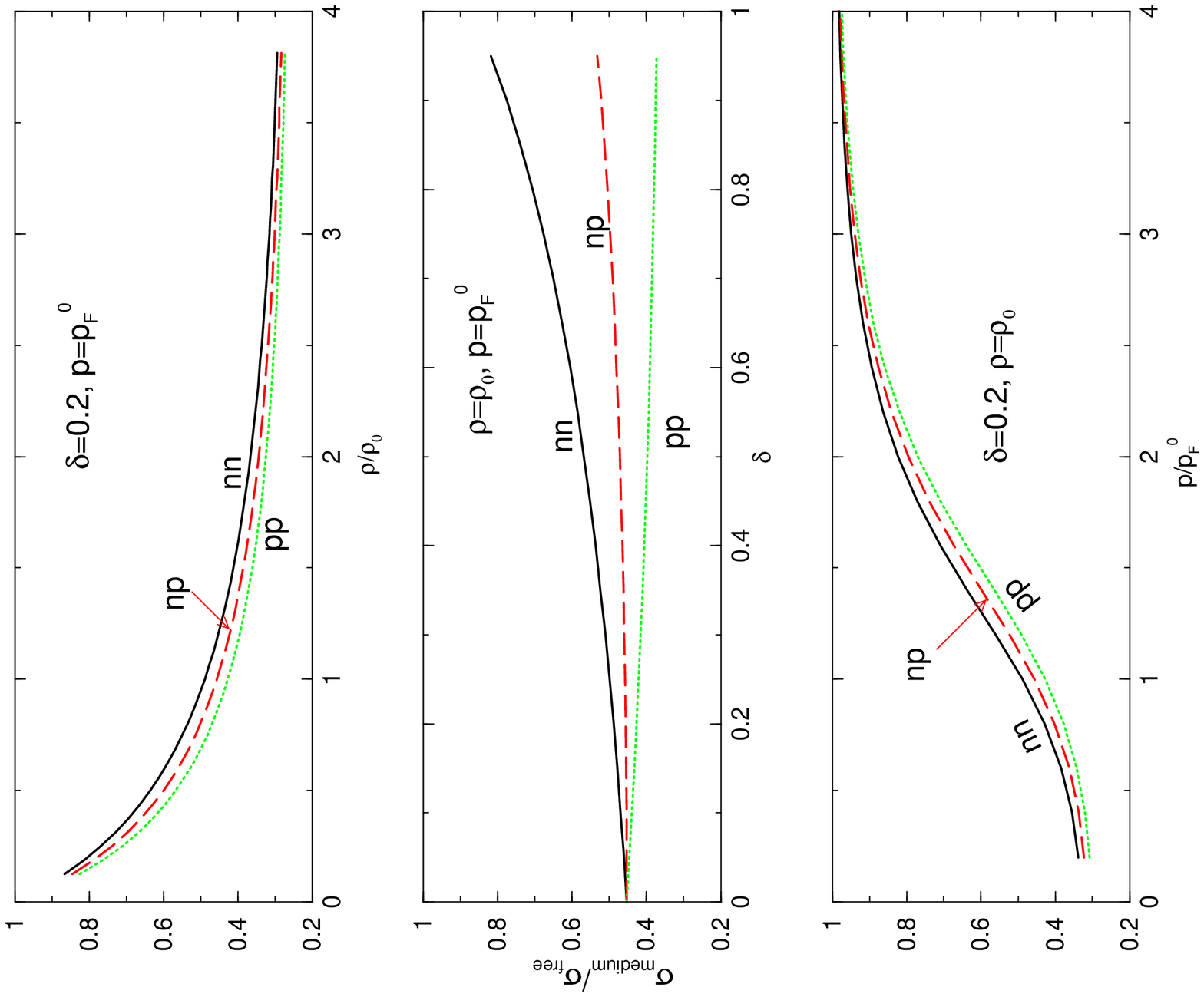}
\caption{Left panels: Neutron and proton effective masses in
asymmetric nuclear matter as functions of density (upper window) and
isospin asymmetry (lower window) from the MDI interaction.
Right panels: Ratio of nucleon-nucleon cross sections in nuclear medium
to their free-space values as a function of density (top window),
isospin asymmetry (middle window), and momentum (bottom window).
Taken from Ref.~\protect\cite{LiChen05}.}
\label{EffMassCrsc}
\end{figure}

In-medium nucleon-nucleon cross sections determine transport properties, such as the stopping power and thermalization rate, of nuclear matter.
In free-space, the neutron-proton cross section is about three times that of proton-proton (neutron-neutron) collisions below the pion production threshold. How does this strong isospin-dependence
of nucleon-nucleon cross sections behave in neutron-rich matter? Answer to this question has important impact on understanding properties of neutron stars and
heavy-ion collisions.  The mean-field potential and the nucleon-nucleon scattering cross sections are
two basic inputs in transport model simulations of heavy ion collisions. In principle, they should be determined self-consistently from the same
interaction. However, due to the complexity of the problem, the nuclear mean-field
potentials and nucleon-nucleon scattering cross sections used in most transport models are obtained separately. In particular, the experimental free-space nucleon-nucleon
scattering cross sections (sometimes with a constant or simple local density-dependent
scalings) are usually applied in some transport model simulations. Physically, it is
expected that the scattering cross sections of {\it pp},  {\it nn} and  {\it np} in
isospin-asymmetric nuclear medium should be modified differently due to the isovector nuclear
effective interactions and the isospin-dependent Pauli blocking. For example, in the IBUU04 transport model \cite{LCK08} and its extended versions, isospin dependent free-space or in-medium
nucleon-nucleon cross sections can be used optionally.  The in-medium cross sections are obtained by extending the effective mass
scaling model in symmetric matter~\cite{Neg81,Pan91,LiGQ94} to isospin asymmetric matter using
the isospin- and momentum-dependent MDI interaction~\cite{LiBA04,Das03,Che05a}.
In the effective mass scaling model, the nucleon-nucleon interaction matrix elements
in the medium are assumed to be identical to that in free-space, and the in-medium
nucleon-nucleon cross sections ($\sigma _{NN}^{\rm medium}$) thus differ
from the free-space ones ($\sigma_{NN}^{\rm free}$) only due to the variation in the
incoming current in the initial state and the density of states in the final state.
Since both factors depend on the effective masses of the colliding nucleon pair, the in-medium nucleon-nucleon cross sections are reduced by the factor
\begin{equation}
R_{\rm medium}\equiv \sigma _{NN}^{\rm medium}/\sigma _{NN}^{\rm free}=(\mu _{NN}^{\ast
}/\mu _{NN})^{2},
\label{xmedium}
\end{equation}%
where $\mu _{NN}$ and $\mu _{NN}^{\ast }$ are, respectively, the free-space and in-medium
reduced masses of the colliding nucleon pair.
It is interesting to mention that the scaling of $\sigma _{NN}^{\rm
medium}/\sigma _{NN}^{\rm free}$ in Eq. (\ref{xmedium}) is found to be
consistent with calculations using the DBHF theory~\cite{Sam05b} for
colliding nucleon pairs with relative momenta less than about $240$ MeV/c at
densities less than about $2\rho _{0}$.

To examine effects of the neutron-proton effective mass splitting on the
isospin-dependence of in-medium nucleon-nucleon cross sections, we take the
MDI interaction as an example \cite{LiChen05}. The left panels of Figure~\ref{EffMassCrsc} display
the effective masses of neutrons and protons in cold asymmetric nuclear matter at their respective Fermi surfaces
as functions of density (upper window) and isospin asymmetry
(lower window) using the MDI interaction \cite{LiChen05}. It is seen that neutrons have a larger effective mass than
protons in neutron-rich matter. Moreover, the neutron-proton effective mass splitting
increases with both the density and isospin asymmetry of the nuclear medium.
Shown in the right panels of Fig.~\ref{EffMassCrsc} is the reduction factor $R_{\rm medium}$
for two colliding nucleons having the same magnitude of momentum $p$ in cold asymmetric
nuclear matter as a function of density (upper window), isospin asymmetry (middle
window), and the nucleon momentum (bottom window). Interestingly, one can see
that not only the nucleon-nucleon cross sections in nuclear medium are reduced
compared with their values in free-space, but the {\it nn} cross sections are
larger than the {\it pp} cross sections in the neutron-rich matter
although their free-space values are identical. In addition, the difference
between the {\it nn} and {\it pp} scattering cross sections becomes larger in
more neutron-rich matter. The larger in-medium {\it nn} cross sections than
{\it pp} ones in neutron-rich matter are completely due to the positive neutron-proton
effective mass splitting in neutron-rich matter with the MDI interaction as shown
in the left panels of Fig.~\ref{EffMassCrsc}. This feature provides a potentially
useful probe of the neutron-proton effective mass splitting in neutron-rich matter and
can be explored in heavy ion collisions induced by neutron-rich nuclei.
It should be pointed out that in transport model simulations of heavy-ion
collisions, the nucleon effective masses used to obtain the in-medium
nucleon-nucleon cross sections in the effective mass scaling model have
to be calculated dynamically in the evolving nuclear medium created during the
collisions~\cite{LiChen05}.

\section{Effects of neutron-proton effective mass splitting in heavy-ion reactions}
\begin{figure}[ht]
\includegraphics[width=6cm,angle=-90]{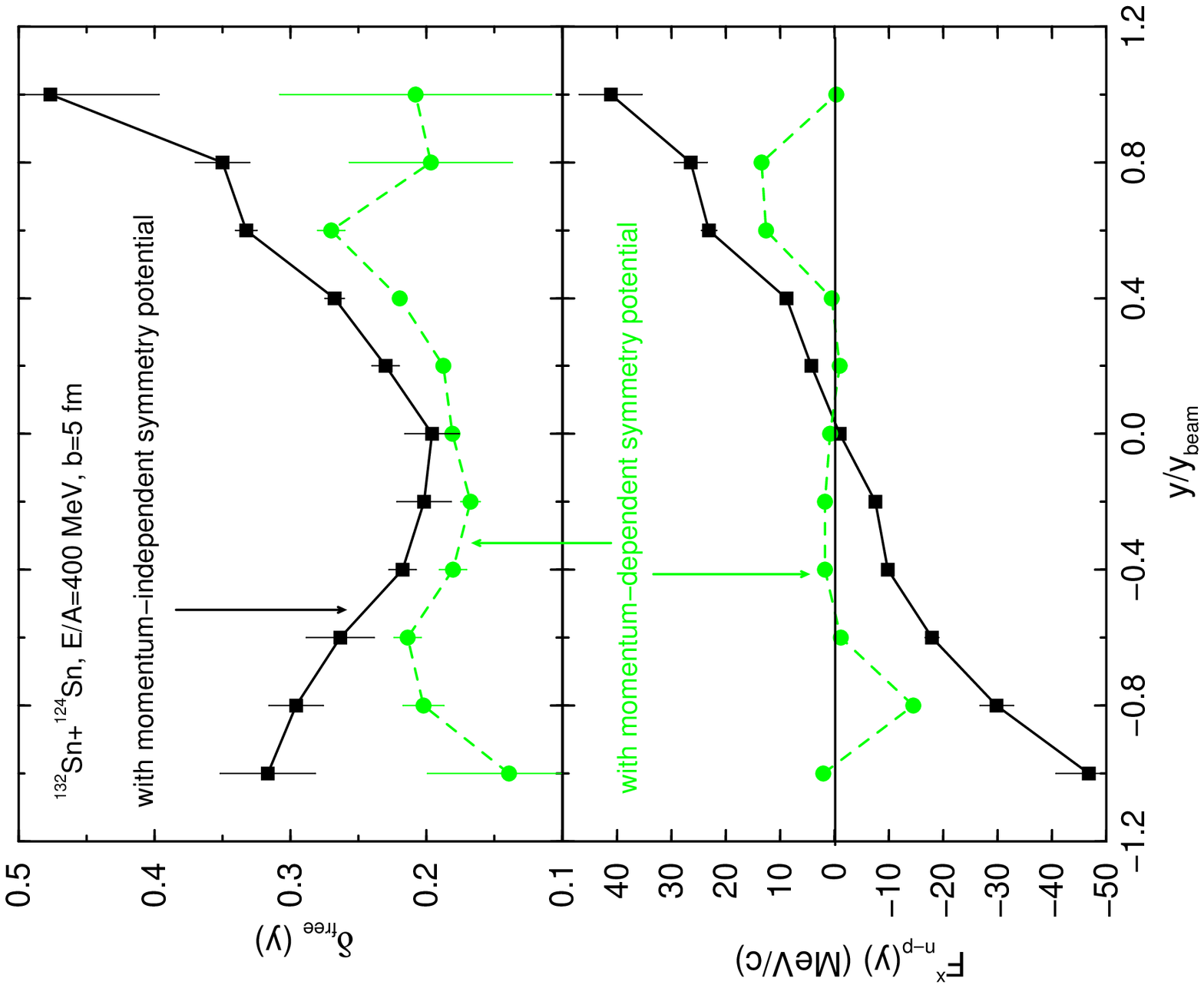}
\includegraphics[width=6cm,angle=-90]{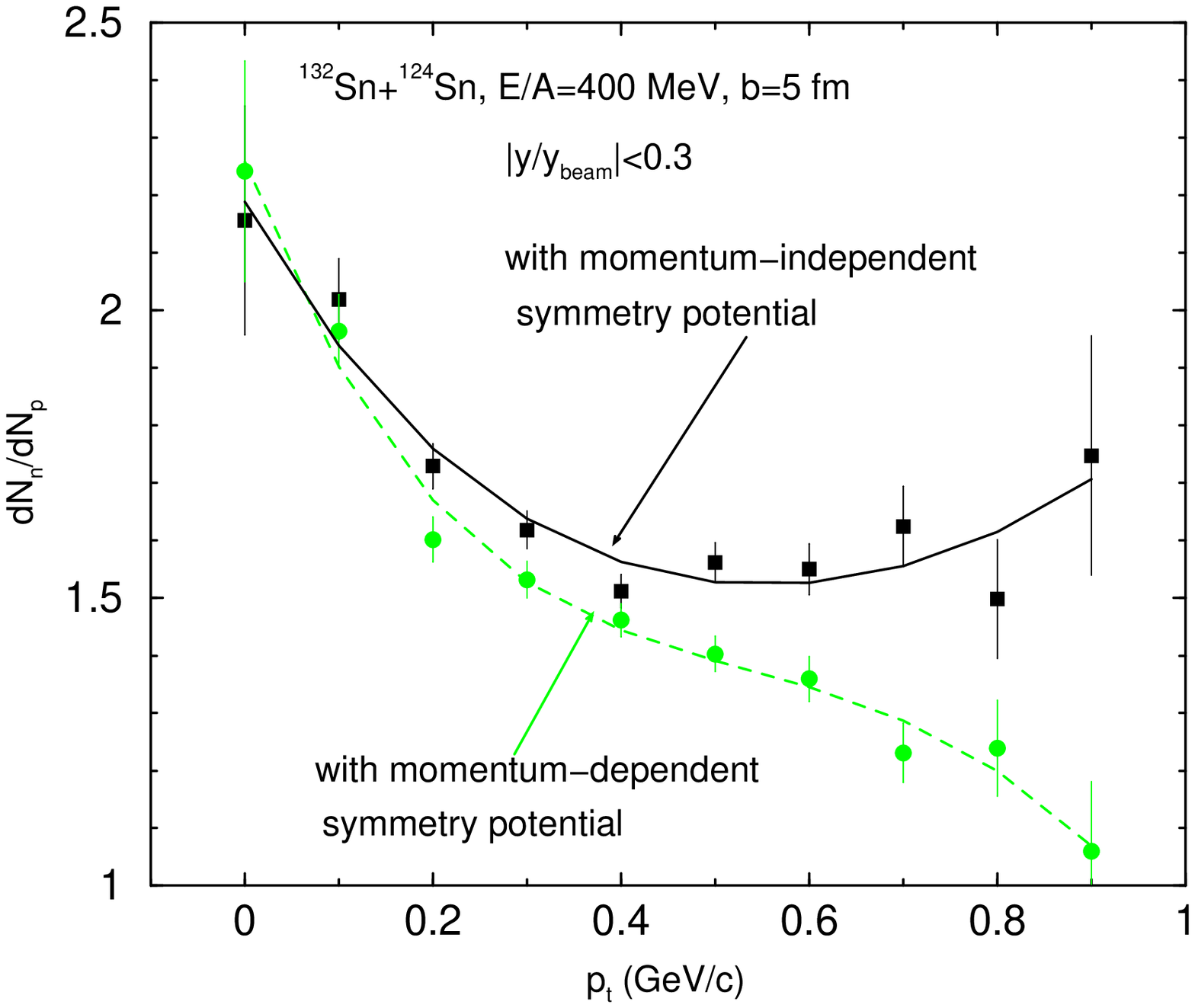}
\caption{{\protect Isospin-asymmetry (Upper left) and neutron-proton differential transverse flow
(lower left) of free nucleons as a function of rapidity. (Right) The ratio of free neutron to proton multiplicity as a function of transverse
momentum at midrapidity. The black solid (green dashed) lines are calculated with the
momentum-independent (-dependent) symmetry potential corresponding to $m^{*}_{n-p}=0$ ($m^{*}_{n-p}>0$). Taken from ref. \protect\cite{LiBA04}.}}
\label{dflow}
\end{figure}
To our best knowledge, the latest neutron-proton effective mass splitting at saturation density extracted from optical model analyses of nucleon-nucleus scattering experiments has not been used in calibrating symmetry potentials predicted by various many-body theories yet. The latter, as we discussed earlier, predict very diverse momentum dependences for the symmetry potential, thus different values for the $m^{*}_{n-p}$ especially at supra-saturation densities. Symmetry potentials used in simulating heavy-ion collisions are consequently very model dependent presently. So far, most of the studies have been focusing on finding experimental observables sensitive to the
density dependence of the symmetry energy and/or the momentum dependence of the symmetry potential, thus the neutron-proton effective mass splitting. Indeed, despite of the often model-dependent predictions, some common features have been found. For instance,  several experimental observables in heavy-ion collisions have been found to be sensitive to the neutron-proton effective mass splitting consistently based on simulations using several different transport models \cite{LiBA04,LiChen05,Riz05,Gio10,Feng12,Zhang14,Xie14}. The identified observables are mainly differential or relative
quantities between neutron-proton or light mirror nuclei where effects of the isoscalar potential are largely canceled out. These observables are also generally insensitive to the in-medium nucleon-nucleon cross sections. As examples, shown in Fig.\ \ref{dflow} are 1) (upper left) the average isospin asymmetry $\delta_{free}(y)$ of free nucleons and 2) (lower left) the neutron-proton differential transverse flow $F^x_{n-p}(y)\equiv\sum_{i=1}^{N(y)}(p^x_iw_i)/N(y)$, where $w_i=1 (-1)$ for neutrons (protons)
and $N(y)$ is the total number of free nucleons at rapidity $y$ from IBUU04 model calculations using the MDI interaction for $^{132}Sn+^{124}Sn$ reactions at a beam energy of 400 MeV/nucleon and an
impact parameter of 5 fm \cite{LiBA04}.  Shown on the right is the n/p ratio of midrapidity nucleons within
$|y_{cms}/y_{beam}|\leq 0.3$ as a function of transverse momentum $p_t$. Its overall decrease at low $p_t$ is due to the Coulomb force.
The two calculations with and without the momentum-dependence of the symmetry potential are done with the same density dependence of the symmetry energy. By construction, they have different neutron-proton effective mass splitting of $m^{*}_{n-p}=0$ (black) and $m^{*}_{n-p}>0$ (green), respectively.
The value of $\delta_{free}(y)$ reflects mainly the degree of isospin fractionation between the free nucleons and the
bound ones at freeze-out. At midrapidity the $\delta_{free}(y)$ values are close to the value expected when a complete isospin
equilibrium is established among all target and projectile nucleons. It is interesting to see that with $m^{*}_{n-p}=0$, the $\delta_{free}(y)$ is significantly higher than with $m^{*}_{n-p}>0$.  Moreover, the
difference tends to increase with rapidity. These features are what one expects from the different strength of the symmetry potential with or without the momentum dependence \cite{LiBA04}. In the case of
$m^{*}_{n-p}=0$, the symmetry potential is a constant but it decreases with momentum in the case of $m^{*}_{n-p}>0$.
It is seen that effects of the neutron-proton effective mass splitting are stronger on energetic nucleons at high rapidity or transverse momenta.
\begin{figure}[ht]
\includegraphics[width=6.6cm,height=5.5cm]{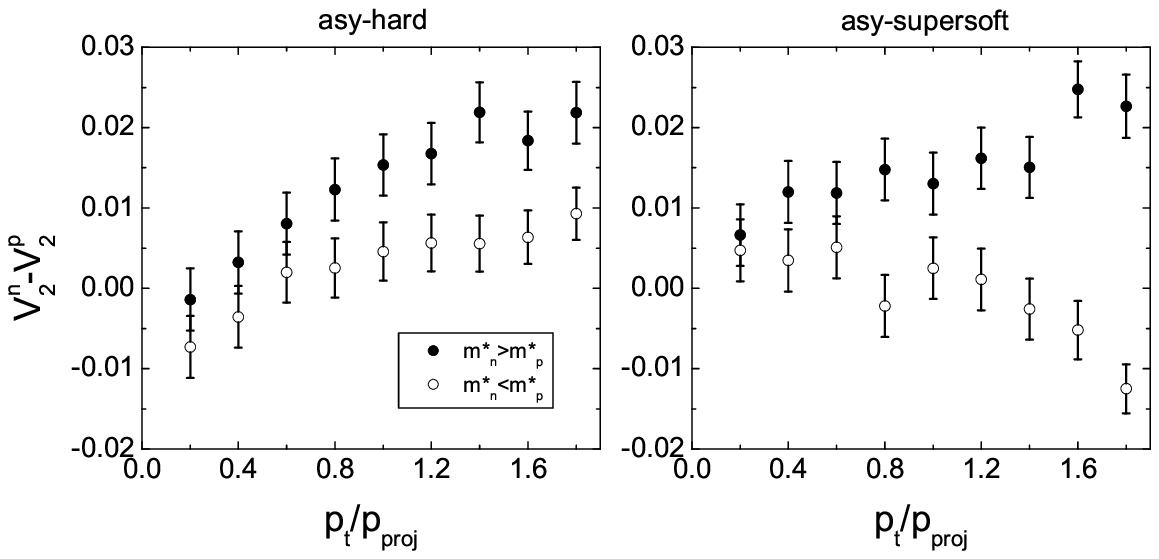}
\includegraphics[width=6cm,height=6.cm]{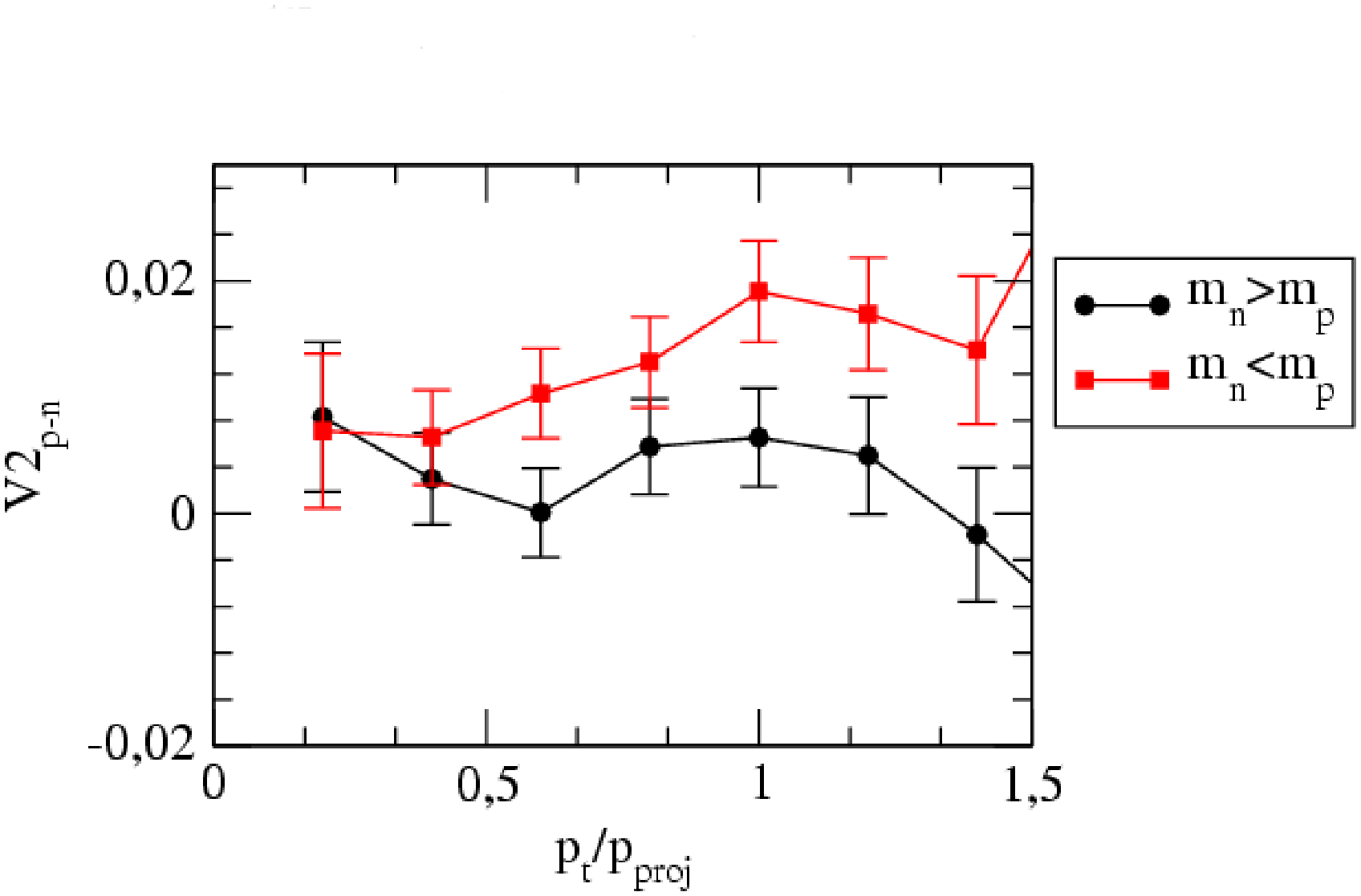}
\caption{A comparison of the difference between neutron and proton elliptic flows with the rapidity bin $|y/y_{proj}|<$0.25 for the different mass splitting in the semi-central $^{197}$Au+$^{197}$Au
collisions using a quantum molecular dynamics model by the Lanzhou group \protect\cite{Feng12}. (Right) Transverse momentum dependence of the difference between proton
and neutron elliptical flow flow $V_2$ at mid-rapidity, $\mid y_0 \mid < 0.3$, in a semi-central reaction Au+Au at 400AMeV using a stochastic mean-field model with a stiff
symmetry energy by the Catania group \protect\cite{Gio10}.}
\label{v2pt}
\end{figure}
Several other observables, such as the $^3$H-$^3$He yield ratio and their difference as a function of energy/rapidity/transverse momentum have also been examined. Conclusions based on these observables are very similar to the ones from using the neutron-proton pairs, see, e.g., ref.~\cite{Che04}. In addition, the relative elliptic flow between neutrons and protons has attracted some special attention. Shown on the left in Fig.\ \ref{v2pt} are the difference between neutron and proton elliptical flows in semi-central $^{197}$Au+$^{197}$Au collisions at 400 MeV/nucleon predicted by a molecular dynamics model \cite{Feng12}. While on the right is the difference between the elliptical flows of protons and neutrons (note the difference in definitions) for the same reaction calculated from a stochastic mean-field model with a stiff symmetry energy \cite{Gio10}. Both calculations show consistently that the difference in elliptical flows of neutrons and protons are sensitive to the neutron-proton effective mass splitting.

Unfortunately, indications on the neutron-proton effective mass splitting from transport model analyses of heavy-ion reactions are currently still unclear~\cite{Kon15}. For instance, analyses of the free neutron/proton double ratio from central $^{124}$Sn+$^{124}$Sn and $^{112}$Sn+$^{112}$Sn collisions at 50 and 120 MeV/nucleon at the NSCL/MSU seem to indicate that protons have a slightly larger effective mass than neutrons based on comparisons with calculations within an improved quantum molecular dynamics model~\cite{Cou14}. However, the preferred Skyrme interactions predict a nucleon symmetry potential that is increasing with nucleon energy/momentum at saturation density in contrast to the findings from optical model analyses of nucleon-nucleus scattering data. Another analysis~\cite{Kon15} of the same data using the IBUU11 transport model found that indeed the assumption of $m^*_n\leq m^*_p$ leads to a higher neutron/proton ratio although the underlying symmetry potential disagrees with the constraints from optical model analyses of nucleon-nucleus scattering data. Moreover, results of the IBUU11 calculations using the $E_{sym}(\rho)$, $U_{sym,1}(\rho,p)$ and $m^*_{n-p}$ all within their current uncertainty ranges still under-predict significantly the NSCL/MSU data. This situation clearly calls for more detailed theoretical studies of the free neutron/proton ratio with different transport models and consider possibly new mechanisms, such as effects of the short-range nucleon-nucleon correlations on the symmetry potential/energy \cite{Hen14,Li15}. On the experimental side, besides improving the precision of measuring neutron spectra which are always a great challenge, reactions with high intensity more neutron-rich beams will certainly enhance effects of the neutron-proton effective mass splitting which is proportional to the isospin asymmetry of the medium.

\section{Summary and Outlook}
In summary, the neutron-proton effective mass splitting in neutron-rich nucleonic matter is a fundamental quantity relevant for understanding many interesting issues in many areas of both nuclear physics and astrophysics. Predictions based on various many-body theories using essentially all available 2-body and 3-body forces are rather diverse mainly because of our poor knowledge about the isospin-dependence of in-medium nuclear effective interactions. Since the neutron-proton effective mass splitting is basically determined by the momentum dependence of the symmetry potential which also affects strongly the density dependence of the symmetry energy, constraints on one of them can help limit the other. Indeed, recent constraints on the density slope of the symmetry energy from many analyses of both terrestrial laboratory data and astrophysical observations indicate clearly that the neutron effective mass is higher than that of protons and their difference grows with the isospin asymmetry of the medium.
Moreover, significant progress in constraining the
neutron-proton effective mass splitting at saturation density was made recently by conducing a rather extensive analyses of huge sets of nucleon-nucleus scattering data using an isospin-dependent optical model.
It was found that $m^{*}_{n-p}=(0.41\pm0.15)\delta$ at saturation density. In our possibly biased opinion, this is so far the most reliable constraint available and can be used as a boundary condition in calibrating calculations by many-body theories. Furthermore, the neutron-proton effective mass splitting has long been known to affect several isospin-sensitive observables in heavy-ion collisions. However, analyses of limited heavy-ion reaction data available using various transport models have not reached a consensus yet regarding the neutron-proton effective mass splitting in neutron-rich matter. It is encouraging to note that the study on the
neutron-proton effective mass splitting and the density-dependence of nuclear symmetry energy are receiving much attention recently both theoretically and experimentally. In particular, experiments using more advanced  neutron and charged particle detectors at various radioactive beam facilities will certainly bring us to the next level of research about the neutron-proton effective mass splitting in neutron-rich matter.
We hope that not only the progress made so far but also the unresolved issues discussed in this brief review will further stimulate investigations on this currently still hotly debated topic relevant for both nuclear physics and astrophysics.

\section*{Acknowledgments}
We thank Bao-Jun Cai, Rong Chen, Champak B. Das, Farrooh Fattoyev, Subal Das Gupta, Charles Gale, Che Ming Ko, Xiao-Hua Li, William G. Newton, Chang Xu, Jun Xu and Wei Zuo for collaborations and discussions on several issues reviewed in this article. This work was supported in part by the U.S. NSF under grant No. PHY-1068022, the CUSTIPEN (China-U.S. Theory Institute for Physics with Exotic Nuclei) under DOE Grant No. DE-FG02-13ER42025, the NNSF of China under Grant Nos. 11320101004, 11275125 and 11135011, the Shanghai Rising-Star Program under grant No. 11QH1401100, the "Shu Guang" project supported by Shanghai Municipal Education Commission and Shanghai Education Development Foundation, the Program for Professor of Special Appointment (Eastern Scholar) at Shanghai Institutions of Higher Learning, the National Basic Research Program of China (973 Program) under Contract Nos. 2015CB856904 and 2013CB834405, and the Science and Technology Commission of Shanghai Municipality (11DZ2260700).


\begin{thebibliography}{10}

\bibitem{Li04} B.A. Li, Phys. Rev. C {\bf 69}, 064602 (2004).
\bibitem{LiBA04} B.A. Li, C.B. Das, S. Das Gupta and C. Gale, Phys. Rev. C {\bf 69}, 011603 (2004);
{\it ibid}, Nucl. Phys. A {\bf 735}, 563 (2004).
\bibitem{LiChen05} B.A. Li and L.W. Chen, Phys. Rev. C{\bf 72}, 064611 (2005).
\bibitem{zuo05} W. Zuo, L.G. Cao, B.A. Li, U. Lombardo, and C.W. Shen, Phys. Rev. C {\bf 72}, 014005 (2005).
\bibitem{XuC10} C. Xu, B.A. Li, and L.W. Chen, Phys. Rev. C \textbf{82}, 054607 (2010).
\bibitem{Rchen} R. Chen, B.J. Cai, L.W. Chen, B.A. Li, X.H. Li, and C. Xu, Phys. Rev. C {\bf 85}, 024305 (2012).
\bibitem{LiXH13} X.H. Li, B.J. Cai, L.W. Chen, R. Chen, B.A. Li and C. Xu, Phys. Lett. B \textbf{721}, 101 (2013).
\bibitem{LiXH14} X.H. Li, W.J. Guo, B.A. Li, L.W. Chen, F.J. Fattoyev and W.G. Newton, Phys. Lett. B 743, 408 (2015).
\bibitem{LiBA13} B.A. Li and X. Han, Phys. Lett. B \textbf{727}, 276 (2013).
\bibitem{Jxu15a} J. Xu, L.W. Chen, and B.A. Li, Phys. Rev. C {\bf 91}, 014611 (2015).
\bibitem{Jeu76} J.P. Jeukenne, A. Lejeune, and C. Mahaux, Phys. Rep. \textbf{25}, 83 (1976).
\bibitem{Mah85} C. Mahaux, P.F. Bortignon, R.A. Broglia and C.H. Dasso, Phys. Rep. 120, 1 (1985).
\bibitem{jamo} M. Jaminon and C. Mahaux, Phys. Rev. C \textbf{40}, 354 (1989).
\bibitem{Fuc05} E.N.E. van Dalen, C. Fuchs and A. Faessler, Phys. Rev. Lett. \textbf{95}, 022302 (2005).
\bibitem{Ron06} Z. Y. Ma, J. Rong, B. Q. Chen, Z. Y. Zhu and H. Q. Song, Phys. Lett. B 604, 170 (2004); W.-H. Long, N. Van Giai and J. Meng, Phys. Lett. B 640, 150 (2006).

\bibitem{Che07} L.W. Chen, C.M. Ko and B.A. Li, Phys. Rev. C \textbf{76}, 054316 (2007).

\bibitem{sjo76} O. Sj\"oberg, Nucl. Phys. A {\bf 265}, 511 (1976).
\bibitem{Neg81} J.W. Negele and K. Yazaki, Phys. Rev. Lett. {\bf 62}, 71 (1981).
\bibitem{Mig57} A.B. Migdal, Sov. Phys. JEPT. \textbf{5}, 333 (1957).
\bibitem{Lut60} J.M. Luttinger, Phys. Rev. {\bf 119}, 1153 (1960).
\bibitem{Czy61} W. Czyz and K. Gottfried, Nucl. Phys. A {\bf 21}, 676 (1961).
\bibitem{Sar80} R. Sartor and C. Mahaux, Phys. Rev. C {\bf 21}, 1546 (1980).
\bibitem{Bla81} J.P. Blaizot and B.L. Friman, Nucl. Phys. A {\bf 372}, 69 (1981).
\bibitem{Kro81} E. Krotscheck, R.A. Smith, and A.D. Jackson, Phys. Lett. B \textbf{104}, 421 (1981).
\bibitem{Jac82} A.D. Jackson, E. Krotscheck, D.E. Meltzer, and R.A.
Smith, Nucl. Phys. A \textbf{386}, 125 (1982).

\bibitem{Mei07} Ulf-G. Mei$\beta$ner, A.M. Rakhimov, A. Wirzba, and U.T. Yakhshiev, Eur. Phys. J. A {\bf 31}, 357 (2007);
{\it ibid}, Eur. Phys. J. A {\bf 32}, 299 (2007); {\it ibid}, Eur. Phys. J. A {\bf 36},37 (2008).
\bibitem{Ste06} G. Steigman, Int. J. Mod. Phys. E \textbf{15}, 1 (2006).
\bibitem{Bur06} A. Burrows, S. Reddy, and T.A. Thompson, Nucl. Phys. A {\bf 777}, 356 (2006).
\bibitem{Nol69} J.A. Nolen and J.P. Schiffer, Ann. Rev. Nucl. Part. Sci. \textbf{19}, 471 (1969).
\bibitem{Wod97} P.J. Woods and C.N. Davids, Ann. Rev. Nucl. Part. Sci. \textbf{47}, 541 (1997).

\bibitem{Cha05} R.J. Charity and L.G. Sobotka, Phys. Rev. C {\bf 71}, 024310 (2005).

\bibitem{Riz05} J. Rizzo, M.Colonna and M. Di Toro, Phys. Rev. C {\bf 72}, 064609 (2005).
\bibitem{Gio10} V. Giordano et al., Phys. Rev. C \textbf{81}, 044611 (2010).
\bibitem{Feng12} Z.Q. Feng, Nucl. Phys. A {\bf 878}, 3 (2012); Phys. Lett. B {\bf 707}, 83 (2012).
\bibitem{Zhang14} Y. Zhang, M.B. Tsang, Z. Li, and H. Liu, Phys. Lett. B {\bf 732}, 186 (2014).
\bibitem{Xie14} W.J. Xie and F.S. Zhang, Phys. Lett. B \textbf{735}, 250 (2014).
\bibitem{Beh11} B. Behera, T.R. Routray, and S.K. Tripathy, J. Phys. G \textbf{38}, 115104 (2011).

\bibitem{Jxu15b} J. Xu, arXiv:1502.02335

\bibitem{Yak01} D.G. Yakovlev, A.D. Kaminker, O.Y. Gnedin, and P. Haensel, Phys. Rep. \textbf{354}, 1 (2001).

\bibitem{Pag06} D. Page and S. Reddy, Ann. Rev. Nucl. Part. Sci. \textbf{56}, 327 (2006).

\bibitem{Bal14} M. Baldo, G. F. Burgio, H.-J. Schulze, and G. Taranto, Phys. Rev. C \textbf{89}, 048801 (2014).

\bibitem{Cou14} D.S. Coupland et al., arXiv:1406.4546

\bibitem{LLL} L.L. Li, Z.H. Li, E.G. Zhao, S.G. Zhou, W. Zuo, A. Bonaccorso, and U. Lombardo, Phys. Rev. C {\bf 80}, 064607 (2009).
\bibitem{Lan62} A. M. Lane, Nucl. Phys. 35, 676 (1962).
\bibitem{Fuc04} E.N.E. van Dalen, C. Fuchs, and A. Faessler, Nucl. Phys. A \textbf{741}, 227 (2004).

\bibitem{Ma04} J. Rong, Z.Y. Ma and N. Van Giai, Phys. Rev. C \textbf{73}, 014614 (2006).

\bibitem{Sam05a} F. Sammarruca, W. Barredo and P. Krastev, Phys. Rev. C \textbf{71}, 064306 (2005).
\bibitem{Ulr97} S. Ulrych, H. M\"{u}ther, Phys. Rev. C \textbf{56}, 1788 (1997).

\bibitem{Bom91} I. Bombaci andU. Lombardo, Phys. Rev. C \textbf{44}, 1892 (1991).

\bibitem{Fri05} S. Fritsch, N. Kaiser, and W. Weise, Nucl. Phys. A \textbf{750}, 259 (2005).

\bibitem{Bar05} V. Baran, M. Colonna, M.Di Toro, M. Zielinska-Pfabe, and
H.H. Wolter, Phys. Rev. C \textbf{72}, 064620 (2005).

\bibitem{Das03} C.B. Das, S. Das Gupta, C. Gale, and B.A. Li, Phys. Rev. C
\textbf{67}, 034611 (2003).
\bibitem{Beh05} B. Behera, T.R. Routray, A. Pradhan, S.K. Patra, and P.K. Sahu,
Nucl., Phys. A \textbf{753}, 367 (2005).
\bibitem{Xu07c} J. Xu, L.W. Chen, B.A. Li, and H.R. Ma,
Phys. Rev. C \textbf{77}, 014302 (2008).

\bibitem{Che05c} L.W. Chen, C.M. Ko, and B.A. Li, Phys. Rev. C \textbf{72}, 064606 (2005).
\bibitem{LiZH06b} Z.H. Li, L.W. Chen, C.M. Ko, B.A. Li, and H.R. Ma, Phys. Rev. C \textbf{74}, 044613 (2006).

\bibitem{LCK08} B.A. Li, L.W. Chen, and C. M. Ko, Phys. Rep. \textbf{464}, 113 (2008).

\bibitem{Gogny} J. Decharge and D. Gogny, Phys. Rev. C \textbf{21}, 1568 (1980); D. Gogny and R. Padjen, Nucl. Phys. A \textbf{293}, 365 (1977).
\bibitem{Ber91} J.F. Berger, M. Girod, and D. Gogny, Comput. Phys. Commun. \textbf{63}, 365 (1991).
\bibitem{Cha08} F. Chappert, M. Girod, and S. Hilaire, Phys. Lett. B \textbf{668}, 420 (2008).
[\bibitem{Gor09} S. Goriely, S. Hilaire, M. Girod, and S. Peru, Phys. Rev. Lett. \textbf{102}, 242501 (2009).

\bibitem{pre} M.A. Preston and R.K. Bhaduri, Structure of the Nucleus, Addison-Wesley, Reading, MA, 1975, p. 191-202.
\bibitem{Xu-tensor} C. Xu and B.A. Li, Phys. Rev. C \textbf{81}, 064612 (2010).
\bibitem{Wang12} Y.N. Wang, J.N. Hu, H. Toki, and H. Shen, Prog. Theor. Phys. \textbf{127}, 739 (2012).

\bibitem{hug} N.M. Hugenholtz and L. van Hove, Physica \textbf{24}, 363 (1958).
\bibitem{bru64} K.A. Brueckner and J. Dabrowski, Phys. Rev. \textbf{134}, B722 (1964).
\bibitem{Dab73} J. Dabrowski and P. Haensel, Phys. Lett. B \textbf{42}, 163 (1972); Phys. Rev. C \textbf{7}, 916 (1973); Can. J. Phys. \textbf{52}, 1768 (1974).
\bibitem{xuli2} C. Xu, B.A. Li, L.W. Chen, and C.M. Ko, Nucl. Phys. A {\bf 865}, 1 (2011).

\bibitem{Hod94} P.E. Hodgson, The Nucleon Optical Model, 1994 (World Scientific).
\bibitem{Cha14} R.J. Charity, W.H. Dickhoff, L.G. Sobotka, and S.J. Waldecker, Eur. Phys. J. A \textbf{50}, 23 (2014); and the Erratum,
Eur. Phys. J. A \textbf{50}, 64 (2014).
\bibitem{Exfor} http:www.nndc.bnl.gov, National Nuclear Data Center, Brookhaven National Laboratory, USA.
\bibitem{Ham90} S. Hama, B.C. Clark, E.D. Cooper, H.S. Sherif, and R.L. Mercer, Phys. Rev. C \textbf{41}, 2737 (1990).
\bibitem{Kon03} A.J. Koning and J.P. Delaroche, Nucl. Phys. A \textbf{713}, 231 (2003).
\bibitem{Jeu91} J.-P. Jeukenne, C. Mahaux, and R. Sartor, Phys. Rev. C \textbf{43}, 2211 (1991).
\bibitem{Rap79} J. Rapaport, V. Kulkarni, and R.W. Finlay, Nucl. Phys. A \textbf{330}, 15 (1979).
\bibitem{Pat76} D.M. Patterson, R.R. Doering, and A. Galonsky, Nucl. Phys. A \textbf{263}, 261 (1976).
\bibitem{Dab64} J. Dabrowski, Phys. Lett. \textbf{8}, 90 (1964).

\bibitem{Pan91} V.R. Pandharipande and S.C. Pieper, Phys. Rev. C \textbf{45}, 791 (1991).
\bibitem{LiGQ94} G.Q. Li and R. Machleidt, Phys. Rev. C \textbf{48}, (1994) 1702; {\it ibid}, C {\bf 49}, 566 (1994).
\bibitem{Che05a} L.W. Chen, C.M. Ko, and B.A. Li, Phys. Rev. Lett. \textbf{94}, 032701 (2005).

\bibitem{Sam05b} F. Sammrruca and P. Krastev, nucl-th/0506081, 2005.

\bibitem{Che04} L.W. Chen, C.M. Ko, and B.A. Li, Phys. Rev. C \textbf{69}, 054606 (2004).

\bibitem{Kon15} H.Y. Kong, Y. Xia, J. Xu, L.W. Chen, B.A. Li, and Y.G. Ma, arXiv:1502.00778, Phys. Rev. C (2015) to be published.

\bibitem{Hen14}O. Hen, B.A. Li, W.J. Guo, L.B. Weinstein, and E. Piasetzky, Phys. Rev. C \textbf{91}, 025803 (2015).

\bibitem{Li15} B.A. Li, W.J. Guo, and Z.Z. Shi, Phys. Rev. C{\bf 91}, 044601 (2015).

\end{thebibliography}

\end{document}